# Trimer Bonding States on the Surface of Transition-metal Dichalcogenide $TaTe_2$

Chen Chen,[1] Heung-Sik Kim,[1] Alemayehu S. Admasu,[1] Sang-Wook Cheong,[1,2] Kristjan Haule,[1] David Vanderbilt,[1] and Weida Wu[1]*

[1]*Department of Physics & Astronomy, Rutgers University, Piscataway, New Jersey 08854, USA*

[2]*Rutgers Center for Emergent Materials, Rutgers University, Piscataway, New Jersey 08854, USA*

## Abstract

We report a comprehensive study on the surface structural and electronic properties of $TaTe_2$ at room temperature. The surface structure was investigated using both low energy electron diffraction intensity versus voltage and density functional theory calculations. The relaxed structures obtained from the two methods are in good agreement, which is very similar to the bulk, maintaining double zigzag trimer chains. The calculated density of states indicates that such structure originates from the trimer bonding states of the Ta $d_{xz}$ and $d_{xy}$ orbitals. This work will further provide new insights towards the understanding of the charge density wave phase transition in $TaTe_2$ at low temperature.

## I. Introduction

Layered transition-metal dichalcogenides (TMDs) have been extensively studied because of their intriguing properties, such as the coexistence or competition of charge density wave (CDW) [1-5] and superconductivity [6-13]. The monolayers and bilayers of these compounds are also promising for future technological applications [14-18]. Their electronic properties are intimately coupled to the distortions in crystal structure. For example, it is shown that various CDW modulations, associated with different polytypes and polymorphs, can dramatically affect the superconducting phase transition temperatures in $TaSe_{2-x}Te_x$ [10,19].

1*T*-$TaTe_2$ belongs to the family of tantalum based 1*T* TMD polymorph $TaX_2$, which exhibits a plethora of states. For instance, the ground state for $TaS_2$ and $TaSe_2$ is Mott insulating, and for $TaTe_2$ it is metallic with commensurate CDW (CCDW) [1,6,20-22]. Superconducting phases can be achieved when S (Se) is partially replaced by Se (Te), which have been shown closely related to the commensurate and incommensurate CDW (ICDW) phases [6,7,20,21]. Their structures also exhibit a variety of modulations. For the undoped compounds, both $TaS_2$ and $TaSe_2$ evolve from ICDW at high temperatures to CCDW at low temperatures with $\sqrt{13} \times \sqrt{13}$ "star of David" superstructure [23-26]. In contrast, $TaTe_2$ has 3×1 double zigzag stripe superstructure at room temperature, and transforms into 3×3 "butterfly" like superstructure around 170 K [22]. It still remains an open question why the more distorted structure at low temperature shows decreased resistance with enhanced magnetic susceptibility [22], which is counterintuitive since CDW-like distortions usually result in reduced density of states at the Fermi level. To understand this behavior, it is beneficial to first proceed from the emergence of 3×1 superstructure in RT phase from the ideal 1*T* structure. Previous scanning tunneling microscopy (STM) measurements on $TaTe_2$ surface at RT suggests that the origin of the stripe structure is the formation of Ta-Ta dimer chains, which is different from its bulk trimer chain structure [6]. This result inspires us to accurately determine the surface atomic structure, because STM probes local density of states at the surface instead of pure atomic positions. One established method is the low energy electron diffraction intensity versus voltage (LEED *I-V*) calculation, which has been proven reliable in quantitatively studying many surface structures including the TMD compounds [27-29]. Furthermore, adapting the atomic structure into the electronic structure calculation can elucidate its underlying mechanism.

In this paper, we first use LEED *I-V* technique to quantitatively determine the surface structure of RT $TaTe_2$. The results show the surface still maintains the double zigzag trimer structure, same as its bulk. Density functional theory (DFT) calculations on the monolayer $TaTe_2$ then reaffirms this structure with the formation of the trimer bonding states of Ta atoms. Such states originate from the partial charge transfer from Te to Ta, maintaining a charge configuration close to $d^{4/3}$ at the Ta sites, which is similar to a previous study [30].

## II. Experimental Techniques

Single crystals of $TaTe_2$ were grown by the chemical vapor transport technique at 900 °C for seven days using iodine as transport agent. The crystals are air-stable gray hexagonal cleavable plates with typical dimensions of 3 × 3 × 0.2 $mm^3$.

$TaTe_2$ samples were cleaved at 300 K in ultra-high vacuum (UHV) environment with base pressure < $1\times10^{-10}$ Torr, and the LEED images were immediately taken using a four-grid LEED optics (OCI BDL800IR) with electron beam energy range from 30 to 450 eV. LEED *I-V* curves were then extrapolated using the methods described in Ref. [31], and smoothed using the Savitzky-Golay method with the third-order polynomial. 47 inequivalent beams with a total energy of $E_{\mathrm{total}}$ = 6600 eV were collected. The structural refinements were performed using a modified version of the symmetrized automated tensor LEED package (SATLEED) [32], with the partial-wave phase shifts calculated using optimized muffin-tin (MT) potential method [33]. Our calculations employed a constant imaginary part of the inner potential of $\mathrm{Im}(V_o)$ = 4.65 eV and total of 9 phase shifts. The Pendry reliability factor $R_P$ was used to characterize the agreement between the experimental and simulated LEED *I-V* curves [27].

For the electronic structure calculations, we employed the Vienna *ab-initio* Simulation Package (VASP), which uses the projector-augmented wave (PAW) basis set [34,35]. A revised Perdew-Burke-Ernzerhof generalized gradient approximation (PBEsol) [36] was employed for an exchange-correlation functional, and a force criterion of 1 meV / Å was adopted for structural optimizations. For the simulation of the RT phase in DFT, which is a zero-temperature tool, the in-plane 3 × 1 periodicity of the RT phase was enforced. The optimized bulk crystal structures, both without and with atomic spin-orbit coupling (SOC), are compared with experimentally reported bulk crystal [22] in Table IV, showing a reasonably good agreement. To explore possible differences between the bulk and surface crystal structures, we employed 5-layer-thick slab geometries for the RT phase with a 25 Å-thick vacuum and a 8 × 4 k-point sampling. We used a 280 eV of plane wave energy cutoff. Effect of SOC was found to be negligible on structural properties. Projected densities of states in Fig. 3 are computed using the single-layer surface slab geometry, which was taken from the optimized five-layer slab calculation.

## III. Results

### A. Surface Superstructure

The structure of bulk crystal is shown in Fig. 1(a), where the cleaving plane is between the Te-Te layers due to their weak bonding. Ideally, the exposed surface of the undistorted 1*T* structure is shown in Fig. 1(b), where the Ta and Te are represented by the blue and red atoms, respectively. At room temperature (RT), previous X-ray analysis indicates that the bulk is already distorted, and the corresponding exposed surface is shown in Fig. 1(c) [22,37]. Here the structural change is due to the emergence of double zigzag trimer chains of the Ta atoms. The top view of the trimer chains in Fig. 1(c) indicates that two Ta atoms (Ta2, blue) move towards one Ta atom in the middle (Ta1, yellow).

The goal of our LEED *I-V* calculation is to clarify the surface structure between the double zigzag trimer and single zigzag dimer chains proposed in Ref. [6]. Under the same symmetry, the structural difference depends only on the bonding distance of different types of Ta atoms. In the bulk, the distance between Ta1 and Ta2 atoms is ~ 3.31 Å and the distance between edge Ta2 atoms is ~ 4.48 Å [22]. The short and long bonding distances are represented by red sticks and blue dashed lines in Fig. 1(c), respectively. Figure 1(d) shows the possible single zigzag dimer chain structure, where the Ta2-Ta2 distance is shorter than the Ta1-Ta2 distance. In Figs. 1(c) and (d), neither the overall surface symmetry nor the lattice parameters of the unit cell are changed. Therefore, to identify the surface structural configuration, one can simply solve the best-fit surface structure and compare the Ta bonding distances.

Figures 1(e) and (f) show the real and schematic LEED patterns of RT $TaTe_2$ surface at 80 eV, consistent with the result in Ref. [6]. For the ideal 1*T* structure, the LEED pattern should consist of integer spots only, which are the red spots in Fig. 1(f). For the RT structure, the surface lattice vectors are 3× along *a* direction and the same along *b* direction, resulting in a 3 × 1 superstructure. Therefore, there are two extra fractional spots between integer spots along *a** direction, represented by the yellow and blue spots in Fig. 1(f). This LEED pattern is a superposition of two 3 × 1 domains along two directions 120° with respect to each other. The major one (domain 1) with relatively higher intensity follows the two 3 × 1 reciprocal lattice vectors *a'** and *b'**, shown by the green and red arrows in Fig. 1(e) and yellow spots in Fig. 1(f). The LEED *I-V* data is collected from domain 1, and the labeling of LEED *I-V* beams is based on 1 × 1 reciprocal lattice vectors *a** and *b**.

During the structural refinement process, atomic positions of the top two $TaTe_2$ monolayers were allowed to relax under the constraint of in-plane opposite movements of Ta and Te pairs with respect to Ta1 atom at the origin of coordinates. Bulk values were used for the rest of the substrate monolayers. This requires 18 atoms in the surface overlayers with 34 independent displacement parameters used in the refinement. The minimum reliability factor achieved is $R_{P,\min} = 0.29$, which is considered acceptable [38-40]. The final optimized structure is defect-free Te terminated, which is expected for a cleaved TMD sample. The stacking of the layers used in the current calculation follows the bulk structure, with the order of "…Ta2-Ta1-Ta2-Ta2-Ta1-Ta2…". The direction of the stacking is nearly perpendicular to the *xy*-plane, with an angle ~ 1.38° off normal caused by the monoclinic crystal structure. Within each monolayer, the stacking is "Te-Ta-Te" and follows the 1*T* structure. We have examined two different stacking sequences by laterally shifting the top monolayer by lattice parameter ±a (*i.e.* ±a'/3). Our calculation shows $R_P = 0.6513$ (6122) when the top layer is shifted by +a (-a), which is much larger than the $R_P$ value with no shifting. We have also examined other terminations with Te deficiency, which yielded much worse $R_P$ larger than 0.66. These results indicate the cleaved surface still follows the bulk termination, which is usually expected for TMD materials.

Figure 2(a) displays 8 of the 47 simulated LEED *I-V* curves from the best-fit structure, in comparison with the experimental data. Cartesian coordinates are employed in the LEED *I-V* simulation. The *x*-axis is parallel to the lattice *a* direction, the *y*-axis is perpendicular to the *x*-axis, and the *z*-axis points out of the plane. The *x*, *y*, and *z* axes are represented by the arrows and the dot in a circle in Fig. 2(b). Deviations of the top monolayer Ta atoms from the bulk coordinates obtained from the LEED *I-V* and the DFT methods are compared in Table I. These numbers agree with each other and their values are very close to zero, suggesting deviations caused by the surface broken symmetry are negligible within the error bars for RT $TaTe_2$. Thus, the surface structure should follow the bulk double zigzag trimer chains with small distortions. The complete set of LEED *I-V* curves and deviations of all Ta and Te atoms are shown in the appendix.

Furthermore, the possibility of dimer chains is examined using LEED *I-V* simulations. Starting from the best-fit structure, the positions of the Ta2 atoms are shifted towards each other, and ended with the dimer chain structure in Fig. 1(d). The $R_P$'s are calculated as a function of the trial structures with various Ta1-Ta2 and Ta2-Ta2 bonding distances, and the result is shown in Fig. 2(c). The separation of the trimer and dimer structures is the boundary between the white and grey shaded area. The minimum $R_P$ for the dimer structure is 0.52, indicating poor agreement between the simulated and experimental *I-V* curves. The estimated error for the $R_P$ is $\Delta R_P = R_{P,\mathrm{min}} \times \sqrt{8|\mathrm{Im}(V_0)|/E_{\mathrm{total}}} = 0.02$ [39]. The red line in Fig. 2(c) indicates the value ($R_{P,\,\mathrm{min}} + \Delta R_P$) = 0.31, which determines the errors of the trial structures through the intersection between the red line and black curve. The Ta1-Ta2 and Ta2-Ta2 bonding distances obtained from the LEED *I-V* and the DFT methods are compared in Table II together with the bulk values, and the difference between these two distances is larger than 1 Å. Small difference (~0.02Å) between the Ta1-Ta2_up and Ta2_down bonding distances from DFT calculations can be noticed in Table II. This can be attributed to the distinction between the two bonds because of the termination of the monoclinic stacking on the surface. Also, it was found from our DFT calculations that the zig-zag dimer chain structure suggested in Ref. [6] does not remain stable. Therefore, the surface of $TaTe_2$ still forms the double zigzag trimer chains, same as the bulk.

Based on the surface trimer structure, the RT STM topography in Ref. [6] can also be well explained. There is no difference in symmetry between the dimer and the trimer structures, so the three types of top layer Te chains in Ref. [6] are still consistent with the trimer structure. They are associated with Te atoms which are on top, outside, and near the boundary of the underlayer Ta trimer chains. Therefore, the close spacing between the chain 2 and 3 in Ref. [6] should be from Te atoms on top of the trimer and near the boundary.

## B. Electronic Structure

The physical mechanism of the trimer chains, both for surface and bulk, can be understood through a dual trimer bonding states of $d_{xz}$ and $d_{xy}$ orbitals, as demonstrated in

Fig. 3. There are three equivalent $t_{2g}$ orbitals in an octahedral environment due to the crystal field splitting, with $d_{xz}$ and $d_{xy}$ orbitals shown in Fig. 3(a). As discussed previously in Ref. [30], in the ideal 1*T* structure consisting of edge-sharing octahedra, two $t_{2g}$ orbitals at nearest-neighboring Ta sites form a strong σ-like overlap, hence resulting in three quasi-one-dimensional chains along the three directions (*x*-*y*, *y*-*z*, *z*-*x*, where the coordinates are defined in Fig. 3(b)) overlaid in a same triangular lattice as shown in Fig. 3(b). Here when the three $t_{2g}$ orbitals from three Ta atoms interact with each other, the resulting molecular orbitals (trimer states) are also of three types, which are illustrated in Fig. 3(c). The energetic order of these orbitals from low to high is: bonding, nonbonding, and antibonding. The key to the double zigzag structure is the formation of trimer states from $d_{xz}$ and $d_{xy}$ orbitals, which are represented in Fig 3(d) as thick yellow and blue lines, respectively.

The calculation of projected densities of states (PDOS) of a monolayer $TaTe_2$ is also consistent with the above argument. Figure 3(e) shows the PDOS for the ideal 1*T* structure (upper panel) and the trimer structure (middle and lower panels). For the 1*T* structure, there is a peak at the Fermi level from the $t_{2g}$ orbitals. When the structure is distorted, the formation of the trimer states splits this peak into bonding, nonbonding, and antibonding states, as labeled above the middle panel in Fig. 3(e). The bonding states are highly occupied below the Fermi level, thus lowering the electronic energy. In a simplified scheme of electron occupancy, one can assume only the bonding states from $d_{xz}$ and $d_{xy}$ orbitals are occupied. Such argument is valid for both Ta1 and Ta2 atoms. Notice that because only the bonding states are occupied, which account for nearly two electrons for a single orbital, there are 4 electrons for $d_{xz}$ and $d_{xy}$ orbitals shared by three Ta atoms. Therefore, the electron count for the Ta is close to $d^{4/3}$, where the extra 1/3 electrons originates from the Te charge transfer [30].

## IV. Discussion and Conclusion

The formation of bonding states not only has successfully explained the structural modulations in several *d*-electron containing systems [30], but also provides important insight in further understanding of the CDW and related phase transition in $TaTe_2$. Note that a structural transition happens at 170 K, where the trimer chain breaks down to butterfly shaped clusters [22]. We propose a similar mechanism with the trimer state formation of the $d_{yz}$ orbital at low temperature. In Fig. 3(c), the density of states of the $d_{yz}$ orbital is large at the Fermi level, especially for the Ta1 atoms. This is similar to the Ta $t_{2g}$ orbitals in the ideal 1*T* structure, which can lower the energy by splitting into multiple trimer states. Accompanied by the structural transition is the additional charge transferred to the Ta sites since extra bonding states are formed. In order to check this hypothesis, investigations on this phase transition as a function of Se doping would be a promising effort. Because of Se's larger electron negativity and weaker interlayer Se-Se interactions, the less charge transfer to Ta could suppress the structural transition, similar to the mechanism of the competition of Ir dimerization and Te-Te interlayer bonding in $IrTe_{2-x}Se_x$ [41-43].

In summary, we have used both LEED *I-V* and DFT methods to quantitatively investigate the room temperature surface structure of $TaTe_2$. The results suggest the surface structure is very similar to the bulk, which maintains the double zigzag trimer chains. Band structure analysis indicates the mechanism of such bonding scheme is the formation of the two trimer states of the Ta $d_{xz}$ and $d_{xy}$ orbitals, associated with partial charge transfer from Te to Ta. This offers a possible solution for the mechanism of the CDW phase transition at low temperature, as well as a pathway to control this transition with external parameters. We expect our research results shown here to have general impact on understanding the CDW phases associated with intra- and inter-layer bonding of the TMD materials.

## Acknowledgement

Primary support for this project came from the National Science Foundation Grant NSF-DMREF-1629059.

## Appendix

## Appendix A: Complete LEED *I-V* curves

Figure 4 shows the complete 47 experimental and simulated LEED *I-V* curves of room temperature $TaTe_2$. The red curves represent the experimental data, and the blue curves represent the simulated curves from the best-fit structure. The inset in each subfigure shows the labeling of each beam and the corresponding $R_P$ factor.

## Appendix B: Coordinates of best-fit structure

Figure 5 presents the structural input used in the LEED *I-V* calculation. Figure 5(a) shows the side view of bulk structure with the labeling of different types of atoms consistent with Table III. Figure 5(b) shows the top view of the $TaTe_2$ layer, where the in-plane unit cell is indicated by the parallelogram. Figures 5(c) ~ (e) show the layer-by-layer top view of the surface monolayer, and each is defined as a composite layer in the LEED *I-V* calculation. The relative positions of different atoms can be implied with respect to the in-plane unit cell in each subfigure. The Cartesian coordinates in Fig. 5(d) is the same as in Fig. 2(b), and is also consistent with Table III.

Table III displays the deviations (with error bars) of all the atoms from their bulk positions obtained from the best-fit structure in LEED *I-V* calculation. The constraint of in-plane opposite movements of Ta and Te pairs with respect to Ta1 atom is applied during the calculation, so the *x* and *y* deviations of atoms in the "down" layers has the negative values of their corresponding "up" layer atoms and no error bars.

## Appendix C: Comparison of experimentally measured and DFT optimized (with and without SOC) bulk 3 × 1 superstructures.

The bulk structure of the RT 3 × 1 phase is optimized in DFT, both with and without SOC, and compared with the experimentally measured structure from Ref. [22]. The results are summarized in Table IV.

## Tables

TABLE I. Displacements of Ta atoms. The bulk values are CDW induced distortions from ideal 1*T* structure obtained from Ref. [22]. The LEED and DFT values are deviations from the bulk positions obtained by LEED *I-V* and DFT.

| Atoms | Bulk (Å) | LEED (Å) | DFT (Å) |
|---|---|---|---|
| $Ta2_{up}$ (x) | 0.386 | 0.04 ± 0.08 | -0.010 |
| $Ta2_{up}$ (y) | 0.219 | 0.01 ± 0.07 | -0.012 |
| $Ta2_{up}$ (z) | 0.076 | -0.03 ± 0.03 | 0.008 |
| Ta1 (x) | 0.000 | 0.00 | 0.001 |
| Ta1 (y) | 0.000 | 0.00 | 0.001 |
| Ta1 (z) | 0.000 | 0.00 ± 0.04 | 0.015 |
| $Ta2_{down}$ (x) | -0.386 | -0.04 | -0.007 |
| $Ta2_{down}$ (y) | -0.219 | -0.01 | -0.008 |
| $Ta2_{down}$ (z) | -0.076 | 0.01 ± 0.04 | -0.003 |

TABLE II. Ta1-Ta2 and Ta2-Ta2 bonding distances for the bulk and surface. The bulk XRD values are obtained from Ref. [22].

| | Bulk XRD (Å) | Bulk DFT (Å) | Surface LEED (Å) | Surface DFT (Å) |
|---|---|---|---|---|
| Ta1-Ta2 | 3.3103(5) | 3.2600 | 3.33 ± 0.07 | 3.27 ($Ta2_{up}$)<br>3.25 ($Ta2_{down}$) |
| Ta2-Ta2 | 4.4828(9) | 4.5219 | 4.4 ± 0.2 | 4.53 |

TABLE III. Complete deviations of Ta and Te atoms from the bulk positions obtained from best-fit structure in LEED *I-V* calculation.

| | | Displacement-LEED (Å) | | | Displacement-DFT (Å) | | |
|---|---|---|---|---|---|---|---|
| | Atom | x | y | z | x | y | z |
| First monolayer | $Te1_{up}$ | 0.028 ± 0.048 | 0.010 ± 0.047 | -0.020 ± 0.019 | -0.010 | -0.012 | 0.012 |
| | $Te2_{up}$ | -0.030 ± 0.045 | 0.031 ± 0.045 | 0.009 ± 0.017 | 0.007 | 0.010 | 0.046 |
| | $Te3_{up}$ | 0.028 ± 0.057 | -0.005 ± 0.053 | 0.027 ± 0.021 | -0.013 | -0.015 | 0.032 |
| | $Ta2_{up}$ | 0.035 ± 0.075 | 0.008 ± 0.073 | -0.034 ± 0.033 | -0.010 | -0.012 | 0.008 |
| | Ta1 | 0.000 | 0.000 | 0.001 ± 0.036 | 0.001 | 0.001 | 0.015 |
| | $Ta2_{down}$ | -0.035 | -0.008 | 0.006 ± 0.037 | -0.007 | -0.008 | -0.003 |
| | $Te3_{down}$ | -0.028 | 0.005 | 0.006 ± 0.025 | -0.009 | -0.010 | 0.022 |
| | $Te2_{down}$ | 0.030 | -0.031 | -0.004 ± 0.035 | 0.001 | 0.002 | 0.015 |
| | $Te1_{down}$ | -0.028 | -0.010 | 0.003 ± 0.037 | -0.002 | -0.002 | 0.009 |
| Second monolayer | $Te1_{up}$ | 0.002 ± 0.098 | -0.014 ± 0.087 | -0.006 ± 0.040 | 0.006 | 0.007 | 0.021 |
| | $Te2_{up}$ | -0.003 ± 0.118 | 0.061 ± 0.123 | -0.003 ± 0.048 | 0.007 | 0.009 | 0.022 |
| | $Te3_{up}$ | -0.026 ± 0.127 | -0.019 ± 0.130 | -0.007 ± 0.049 | 0.002 | 0.002 | 0.024 |
| | $Ta2_{up}$ | 0.047 ± 0.142 | 0.003 ± 0.143 | -0.002 ± 0.092 | 0.004 | 0.005 | 0.016 |
| | Ta1 | 0.000 | 0.000 | -0.030 ± 0.083 | 0.006 | 0.008 | 0.016 |
| | $Ta2_{down}$ | -0.047 | -0.003 | -0.010 ± 0.079 | 0.003 | 0.004 | 0.019 |
| | $Te3_{down}$ | 0.026 | 0.019 | -0.015 ± 0.109 | 0.004 | 0.005 | 0.016 |
| | $Te2_{down}$ | 0.003 | -0.061 | -0.014 ± 0.198 | 0.001 | 0.001 | 0.015 |
| | $Te1_{down}$ | -0.002 | 0.014 | -0.026 ± 0.103 | 0.005 | 0.006 | 0.013 |

TABLE IV. Comparison between the experimentally reported (Ref. [22]) and optimized (with and without SOC) bulk crystal structures for the $3 \times 1$ RT-$TaTe_2$.

| | | Experimental | | | DFT-optimized | | | DFT-optimized | | |
|---|---|---|---|---|---|---|---|---|---|---|
| | | Ref. [22], @ T = 298K | | | PBEsol, without SOC | | | PBEsol, with SOC | | |
| | | ($a,b,c$) = (14.784, 3.6372, 9.345) Å | | | ($a,b,c$) = (14.583, 3.602, 9.336) Å | | | ($a,b,c$) = (14.617, 3.604, 9.349) Å | | |
| | | $\beta$ = 110.93 | | | $\beta$ = 110.936 | | | $\beta$ = 111.254 | | |
| Atom | Site | x | y | z | x | y | z | x | y | z |
| Ta1 | 2a | 0 | 0 | 0 | 0 | 0 | 0 | 0 | 0 | 0 |
| Ta2 | 4i | 0.8602 | 1/2 | 0.7091 | 0.8631 | 1/2 | 0.7119 | 0.8629 | 1/2 | 0.7122 |
| Te1 | 4i | 0.0055 | 0 | 0.3093 | 0.0012 | 0 | 0.3082 | 0.0018 | 0 | 0.3085 |
| Te2 | 4i | 0.8515 | 1/2 | 0.9890 | 0.8494 | 1/2 | 0.9886 | 0.8495 | 1/2 | 0.9887 |
| Te3 | 4i | 0.7966 | 1/2 | 0.3775 | 0.7966 | 1/2 | 0.3820 | 0.7966 | 1/2 | 0.3822 |

## Figures

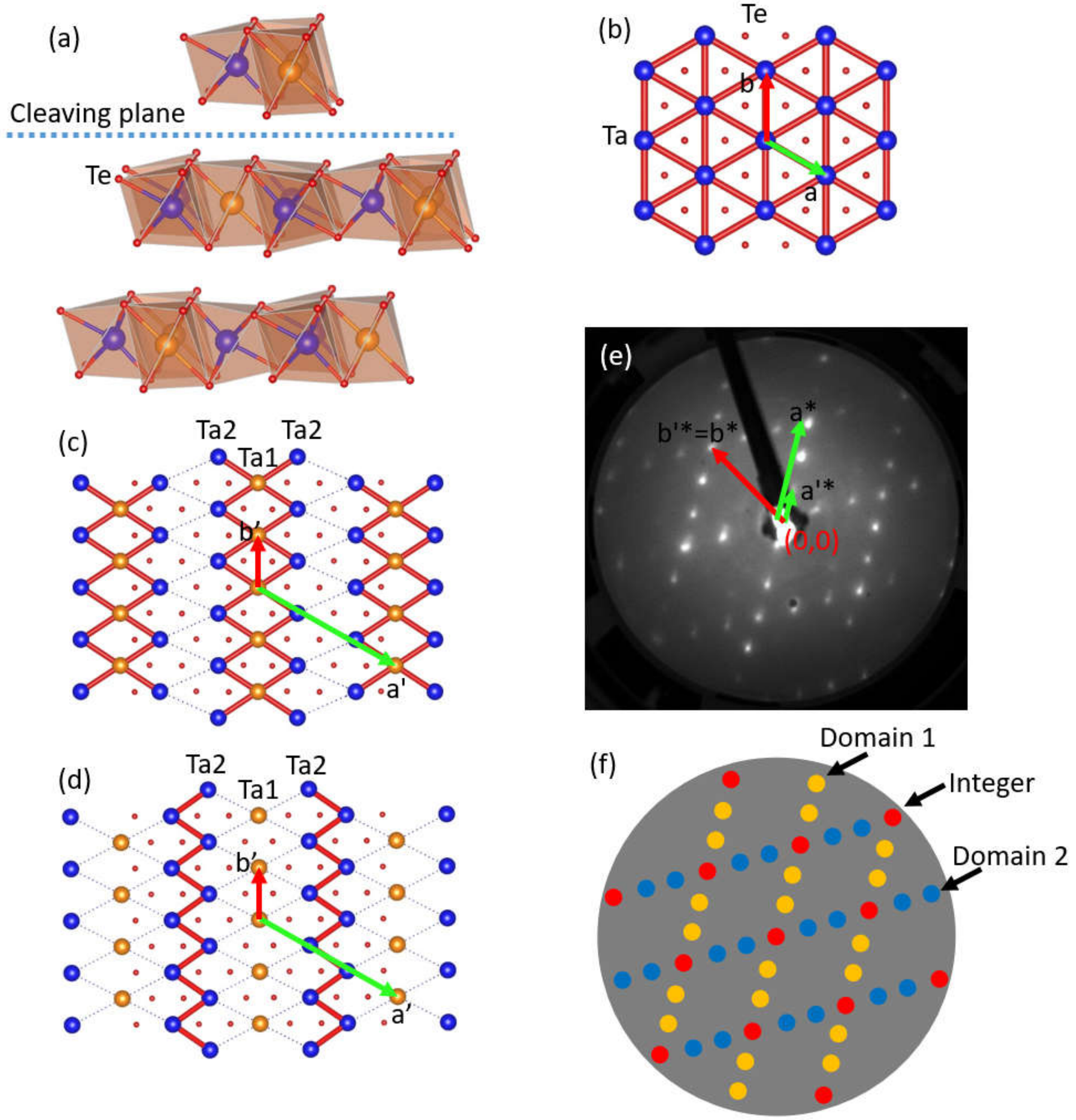

FIG. 1 (color online). (a) Bulk crystal structure of $TaTe_2$. Ta atoms are labeled yellow and blue, and Te atoms are labeled red. Samples are cleaved between the two Te layers. (b) Top view of the undistorted ideal 1*T* structure. (c) Top view of the double zigzag trimer chain structure. Ta1 and Ta2 atoms are labeled yellow and blue, respectively. The double zigzag trimer structure is formed with both Ta2 atoms moving towards Ta1 atoms. The Ta1-Ta2 bondings are indicated by red sticks. The 2D lattice vectors *a* and *b* are labeled by the green and red arrows, forming a 3×1 superstructure. (d) Top view of the single zigzag dimer chain structure. The Ta2-Ta2 bondings are indicated by red sticks. The unit cell has the same lattice vectors, which are also labeled by the green and red arrows. (e) LEED pattern of freshly cleaved $TaTe_2$ surface at 300 K and 80 eV. The reciprocal lattice vectors of domain 1, $a^*$ and $b^*$, are labeled by the green and red arrows. (f) Schematic LEED pattern from a 3×1 superstructure with two domains. The red spots are integer diffraction spots from an ideal 1*T* structure, and the yellow and blue spots are fractional diffraction spots from domain 1 and domain 2, respectively.

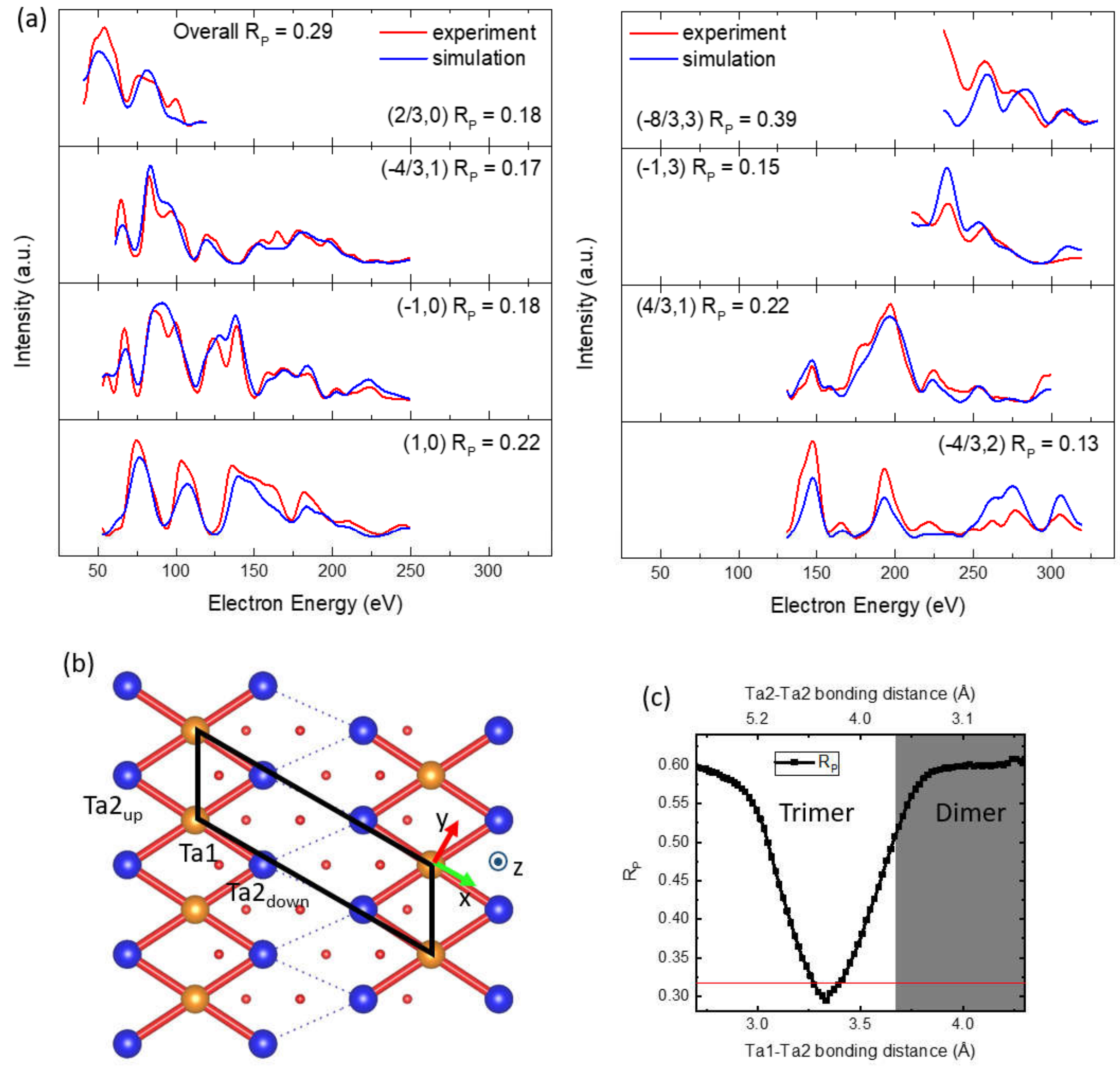

FIG. 2 (color online). (a) Comparison between experimental and simulated LEED *I-V* curves for the best-fit structure. Results for 8 of the total 47 inequivalent beams are represented here. The labeling of each beam is based on the reciprocal lattice vectors $a^*$ and $b^*$ in Fig. 1(e). The overall Pendry reliability factor is $R_P = 0.29 \pm 0.02$. (b) Top view of the $TaTe_2$ surface. The Cartesian coordinates used in LEED *I-V* calculation are labeled by the green and red arrows, and dot in a circle. (c) Dependence of $R_P$ values on the Ta-Ta bonding distances as the structure changes from trimer to dimer.

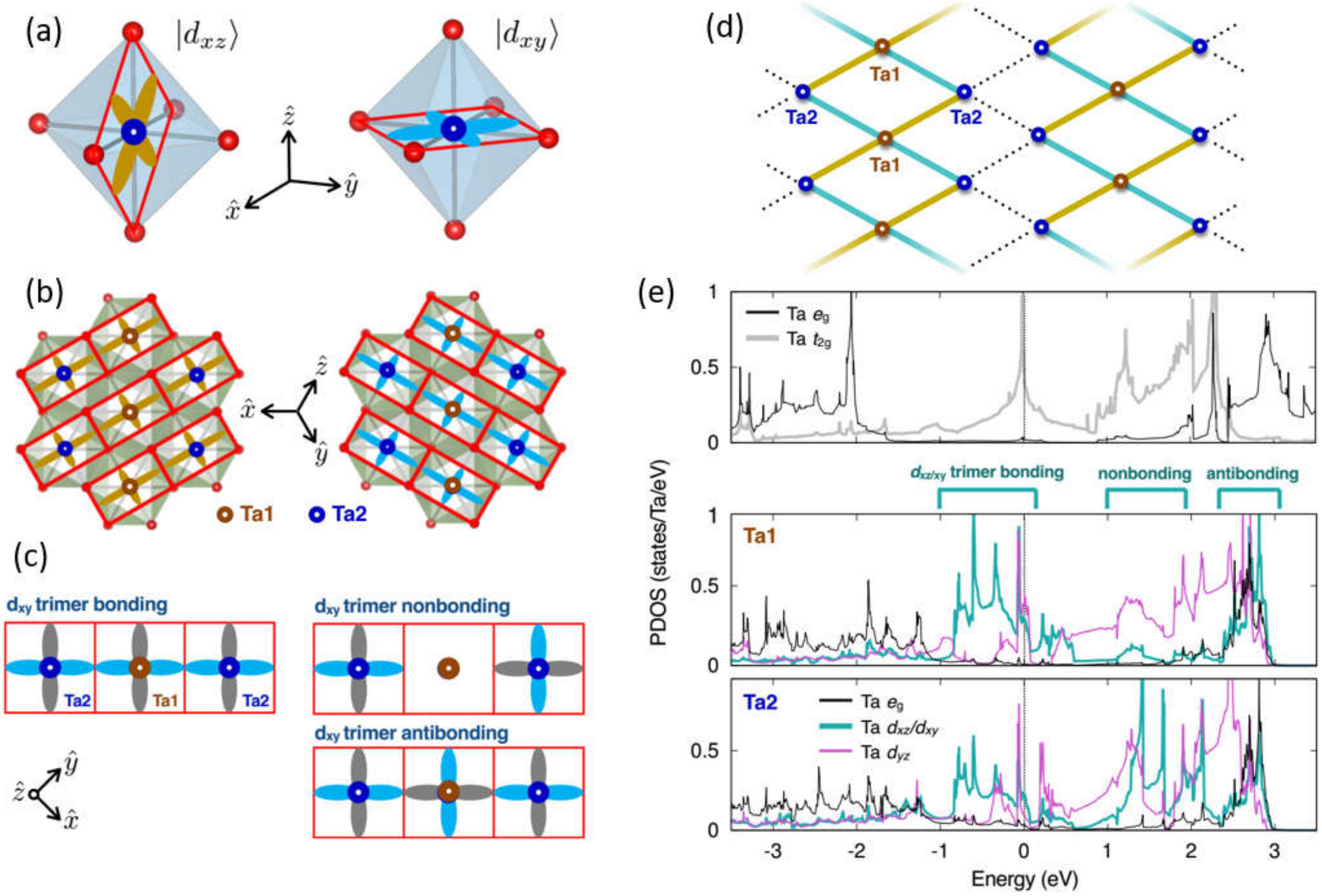

FIG. 3 (color online). (a) Schematic illustration of $d_{xz}$ and $d_{xy}$ orbitals in an octahedral environment. The red squares indicate the planes formed by corresponding coordinates. (b) $d_{xz}$ and $d_{xy}$ orbitals at Ta sites, participating in strong σ-overlaps along *xz* and *xy* directions respectively. The red rectangles are consistent with the red squares in (a). (c) Schematic illustration of bonding, nonbonding, and antibonding states. (d) Schematic illustration of dual trimerization of $d_{xz}$ and $d_{xy}$ chains, depicted as thick dark yellow and blue lines respectively, resulting in the formation of the diamond-shaped chains along the *yz* direction as shown in the figure. (e) Projected densities of states (PDOS) for the hypothetical $C_3$-symmetric structure (upper panel) and for the room temperature structure (middle and lower panels). Note that, due to the trimerization, $d_{xz}$ (or $d_{xy}$) orbitals split into bonding, nonbonding, and antibonding states.

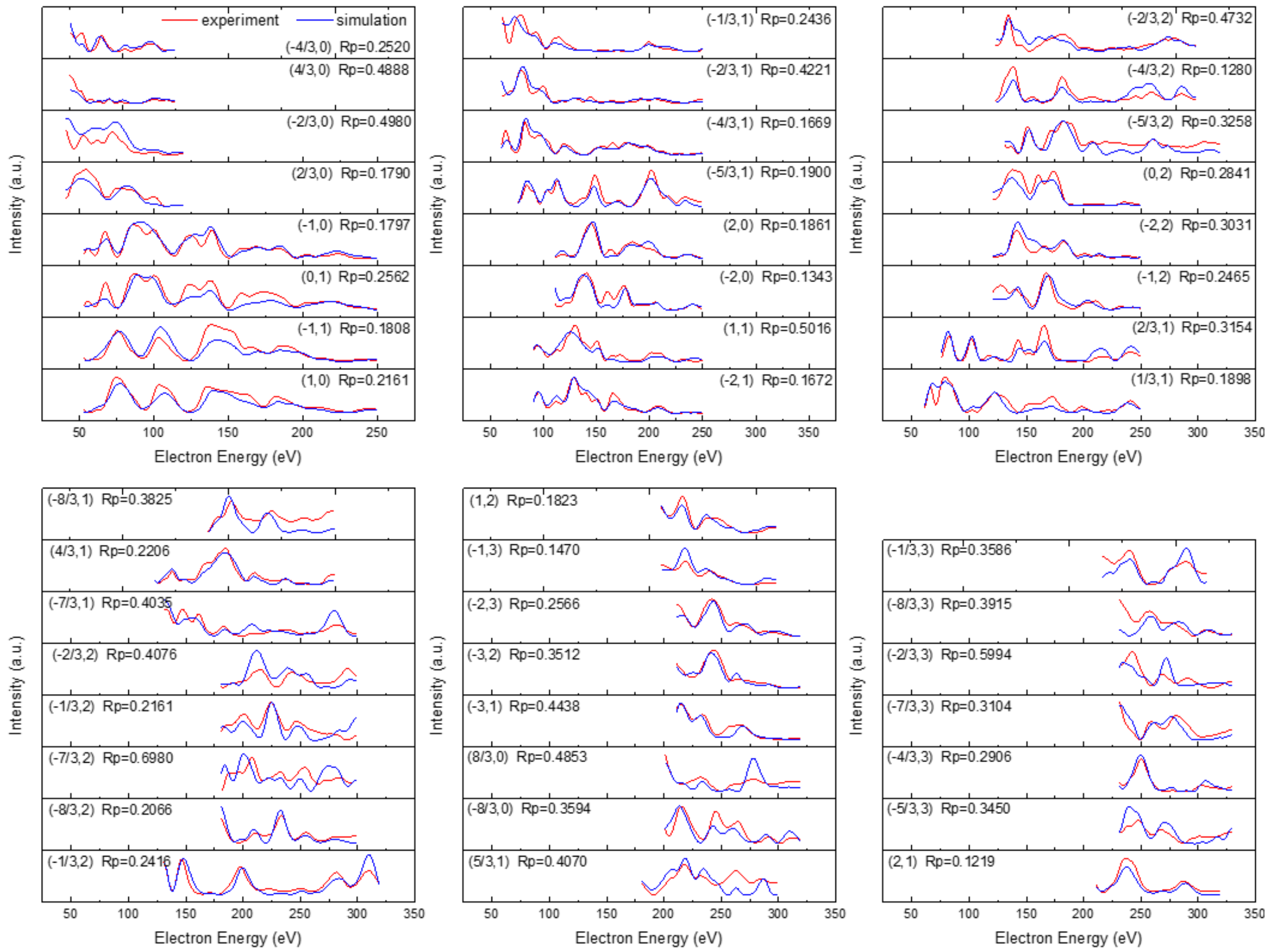

FIG. 4 (color online). Complete comparison between experimental and simulated LEED *I-V* curves for the total 47 inequivalent beams.

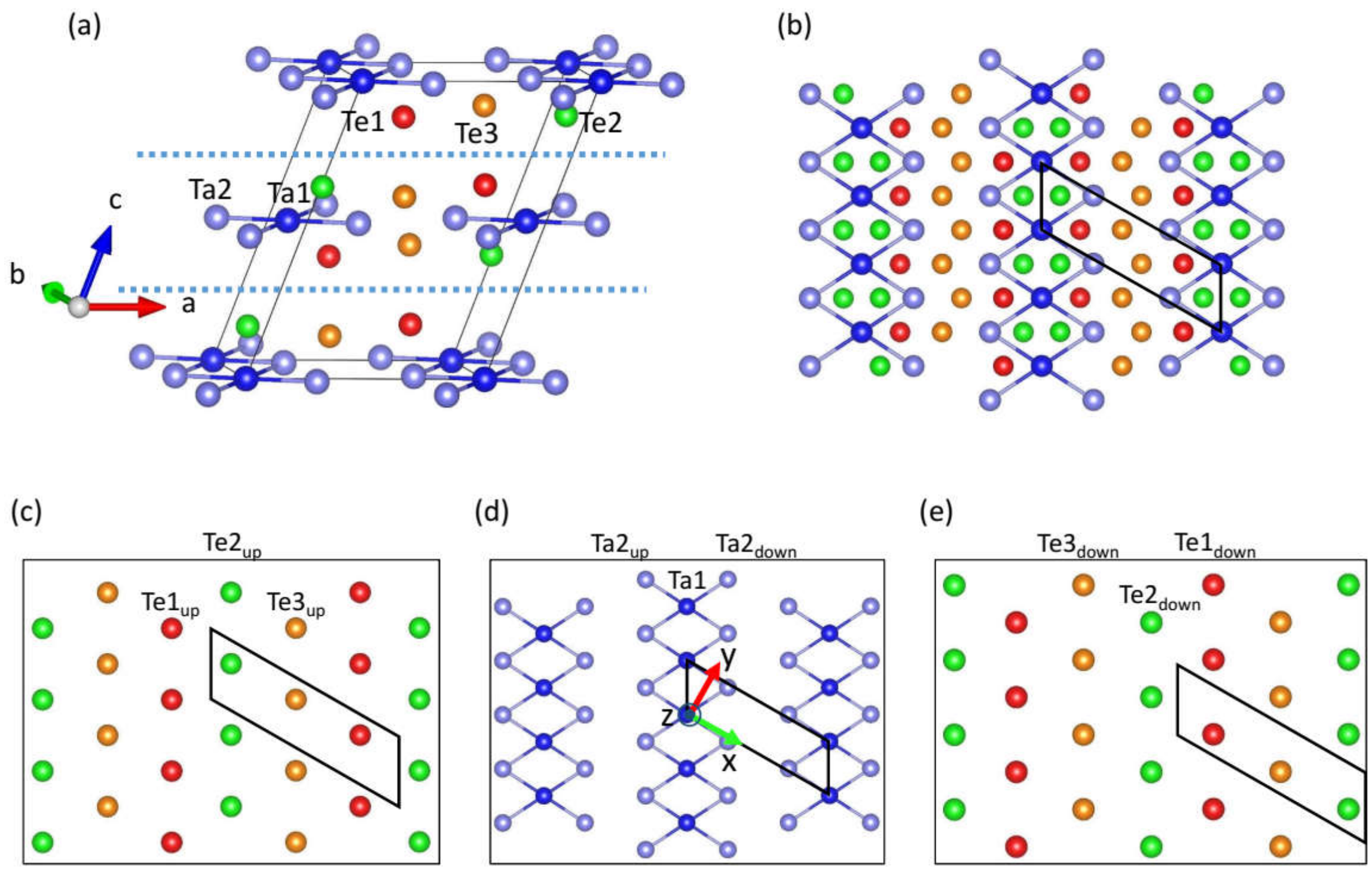

FIG. 5 (color online). (a) Side view of $TaTe_2$ crystal structure. Different types of atoms are as labeled. The cleaving planes are indicated by the dashed lines. (b) Top view of the $TaTe_2$ crystal structure. The black parallelogram indicates the in-plane unit cell. (c) ~ (e) Top view of each composite layer. The top Te layer, Ta layer, and the bottom Te layer are shown by (c), (d), and (e), respectively. The black parallelogram in each subfigure indicates the in-plane unit cell. The Cartesian coordinates used in the calculation are labeled by the green and red arrows, and dot in a circle ($x$, $y$ and $z$) in (b).